\documentclass[prl]{revtex4}
\usepackage{bm}
\usepackage{epsfig}
\usepackage{amsmath}
\usepackage[pdftex,ps2pdf,bookmarks=true,bookmarksnumbered=true,breaklinks=true,hypertexnames=false,linkbordercolor={0 0 1},pdfborder={0 0 112.0}]{hyperref}
\begin{document}
\title{Predissociation of Diatomic Molecules: An Analytically Solvable Model}
\author{Aniruddha Chakraborty \\
School of Basic Sciences, Indian Institute of Technology Mandi,\\
Mandi, Himachal Pradesh, 750001, India.}
\date{\today}
\begin{abstract}
\noindent A model consisting of a Harmonic Oscillator well and a linear potential, coupled by Dirac delta function, is solved. We find the exact analytical expressions for Green's function for this problem. This Green's functions are used to calculate the Raman excitation profile and electronic absorption spectrum.
\end{abstract}
\maketitle

\section{Introduction}
\noindent Nonadiabatic transition due to potential curve crossing is one of the most important mechanisms to
effectively induce electronic transitions in collisions \cite{Naka1,R1,R2,R3,R4,R5,AniBook,AniThesis,Ani1,Ani2,Ani3,Ani4,Ani5,AniRev}. Two state curve crossing can be classified into the following two cases according to the crossing scheme: (1) Landau-Zener (L.Z) case, in which the two diabatic potential curves have the same signs for the slopes and (2) non-adiabatic tunnelling (N.T) case, in which the diabatic curves have the opposite sign for slopes. There is also a non-crossing non-adiabatic transition called the Rosen-Zener-Demkov type \cite{Naka1,Naka2}, in which two adiabatic
potentials are in near resonance at large $R$. The theory of non-adiabatic transitions dates back to $1932$, when the
pioneering works for curve-crossing and non-crossing were published by Landau \cite{Landau}, Zener \cite{Zener} and
Stueckelberg \cite{Stueckelberg} and by Rosen and Zener \cite{Rosen} respectively. Since then numerous papers by many
authors have been devoted to these subjects, especially to curve crossing problems\cite{Naka1,Naka2}. In our earlier paper we have proposed an exactly solvable model for the two state curve crossing problem which assumes the coupling to be a Dirac delta function \cite{Ani1}. This model is used to calculate the effect of curve crossing on electronic absorption spectrum and on Resonance Raman excitation profile for the case of harmonic potentials \cite{Ani2}. We have later generalized our model to deal with general multi-channel curve crossing problem too \cite{Ani3}. Even very recently our model ia extended to deal with nonadiabatic tunneling in an ideal one dimensional semi-infinite periodic potential systems \cite{Ani4}. Our work is in progress to deal with nonadiabatic tunneling in an ideal one dimensional finite periodic potential systems \cite{Ani5}. In the present paper, we describe the predissociation of a diatomic molecule as a decaying quasi-bound vibrational state, to allow for an intersection of two electronic states. The model we propose to solve consists of two diabatic potential curves, approximated by a harmonic oscillator and a linear potential, coupled through a delta function at the crossing point. This model mimics the case in which we have an excited electronic state with a minimum, crossing a purely repulsive state. The advantage of this model is that it can be solved analytically.

\section{The model}
\noindent We consider two diabatic curves, crossing each other. There is a coupling between the two curves, which causes
transitions from one curve to another. This transition would occur in the vicinity of the crossing point. In particular, it will occur in a narrow range of $x$, given by
\begin{equation}
\label{1}\left|V_1(x)-V_2(x)\right|\simeq \left|V(x_c)\right|.
\end{equation}
where $x$ denotes the nuclear coordinate and $x_c$ is the crossing point. $V_1$ and $V_2$ are the diabatic curves and $V$ represent the coupling between them. In reality the transition between $V_{1}(x)$ and $V_{2}(x)$ occur most effectively at the crossing, because the necessary energy transfer between the electronic and nuclear degrees of freedom is minimum there. Therefore it is interesting to analyze a model, where coupling is localized in space near $x_c$ rather than using a model where coupling is same everywhere (i.e. constant coupling). Thus we put
\begin{equation}
\label{2}V(x)=K_0\delta (x-x_c),
\end{equation}
here $K_0$ is a constant. This model has the advantage that it can
be exactly solved \cite{AniThesis,AniBook,Ani1,Ani2,Ani3,Ani4,Ani5,AniRev}.

\section{Exact analytical solution}
\noindent We start with a particle moving on any of the two
diabatic curves. The problem is to calculate the probability that
the particle will still be on any one of the diabatic curves after
a time $t$. We write the probability amplitude as
\begin{equation}
\label{3}\Psi (x,t)=\left(
\begin{array}{c}
\psi _1(x,t) \\
\psi _2(x,t)
\end{array}
\right) ,
\end{equation}
where $\psi _1(x,t)$ and $\psi _2(x,t)$ are the probability amplitude for
the two states. $\Psi (x,t)$ obey the time dependent Schr$\stackrel{..}{o}$
dinger equation (we take $\hbar =1$ here and in subsequent calculations)
\begin{equation}
\label{4}i\frac{\partial \Psi (x,t)}{\partial t}=H\Psi (x,t).
\end{equation}
$H$ is defined by
\begin{equation}
\label{5}H=\left(
\begin{array}{cc}
H_1(x) & V(x) \\
V(x) & H_2(x)
\end{array}
\right) ,
\end{equation}
where $H_i(x)$ is
\begin{equation}
\label{6}H_i(x)=-\frac 1{2m}\frac{\partial ^2}{\partial x^2}+V_i(x).
\end{equation}
We find it convenient to define the half Fourier Transform $\overline{\Psi }%
(\omega )$ of $\Psi (t)$ by
\begin{equation}
\label{9}\overline{\Psi }(\omega )=\int_0^\infty \Psi (t)e^{i\omega t}dt.
\end{equation}
Half Fourier transformation of Eq. (\ref{4}) leads to
\begin{equation}
\label{11}\overline{\Psi }(\omega )=iG(\omega )\Psi (0),
\end{equation}
where $G(\omega )$ is defined by 
\begin{equation}
(\omega -H) G(\omega )=I. 
\end{equation}
In the position representation, the above equation may be written as
\begin{equation}
\label{12}\overline{\Psi }(x,\omega )=i\int_{-\infty }^\infty G(x,x_0;\omega
)\overline{\Psi }(x_0,\omega )dx_0,
\end{equation}
where $G(x,x_0;\omega )$ is
\begin{equation}
\label{13}G(x,x_0;\omega )=\langle x|(\omega -H)^{-1}|x_0\rangle .
\end{equation}
Writing
\begin{equation}
\label{14}G(x,x_0;\omega )=\left(
\begin{array}{cc}
G_{11}^{}(x,x_0;\omega ) & G_{12}^{}(x,x_0;\omega ) \\
G_{21}^{}(x,x_0;\omega ) & G_{22}^{}(x,x_0;\omega )
\end{array}
\right)
\begin{array}{cc}
&  \\
&
\end{array}
\end{equation}
and using the partitioning technique \cite{Lowdin} we can write
\begin{equation}
\label{15}G_{11}^{}(x,x_0;\omega )=\langle x|[\omega -H_1-V(\omega
-H_2)^{-1}V]^{-1}|x_0\rangle.
\end{equation}
The above equation is true for any general $V$. This expression
simplify considerably if $V$ is a delta function located at $x_c$ \cite{AniBook,AniThesis,Ani1,Ani2,Ani3,Ani4,Ani5,AniRev}.
\begin{equation}
\label{22} G_{11}(x,x_0;\omega )=
G_1^0(x,x_0;\omega)+\frac{K_0^2G_1^0(x,x_c;\omega)G_2^0(x_c,x_c;\omega
)G_1^0(x_c,x_0;\omega )}{ 1-K_0^2G_1^0(x_c,x_c;\omega
)G_2^0(x_c,x_c;\omega )},
\end{equation}
where 
\begin{equation}
\label{17}G_{i}^0(x,x_0;\omega )=\langle x|(\omega -H_{i}^{})^{-1}|x_0\rangle ,
\end{equation}
and corresponds to propagation of the particle starting at $x_0$
on the second diabatic curve, in the absence of coupling to the
first diabatic curve. Using the same procedure one can get \cite{AniBook,AniThesis,Ani1,Ani2,Ani3,Ani4,Ani5,AniRev}
\begin{equation}
\label{23}
\begin{array}{c}
G_{12}^{}(x,x_0;\omega )=\frac{K_0G_1^0(x,x_c;\omega )G_2^0(x_c,x_0;\omega )%
}{1-K_0^2G_1^0(x_c,x_c;\omega )G_2^0(x_c,x_c;\omega )}.
\end{array}
\end{equation}
Similarly one can derive expressions for $G_{22}^{}(x,x_0;\omega )$ and $%
G_{21}^{}(x,x_0;\omega )$. Using these expressions for the Green's
function in Eq. (\ref{11}) we can calculate $\overline{\Psi
}(\omega )$ explicitly.
\newline
The expressions that we have obtained for $\overline{\Psi }(\omega
)$ are quite general and are valid for any $V_1(x)$ and $V_2(x)$ \cite{AniBook,AniThesis,Ani1,Ani2,Ani3,Ani4,Ani5,AniRev}.
However, their utility is limited by the fact that one must know
$G_1^0(x,x_0;\omega )$ and $G_2^0(x,x_0;\omega )$.
\begin{figure} \centering
\epsfig{file=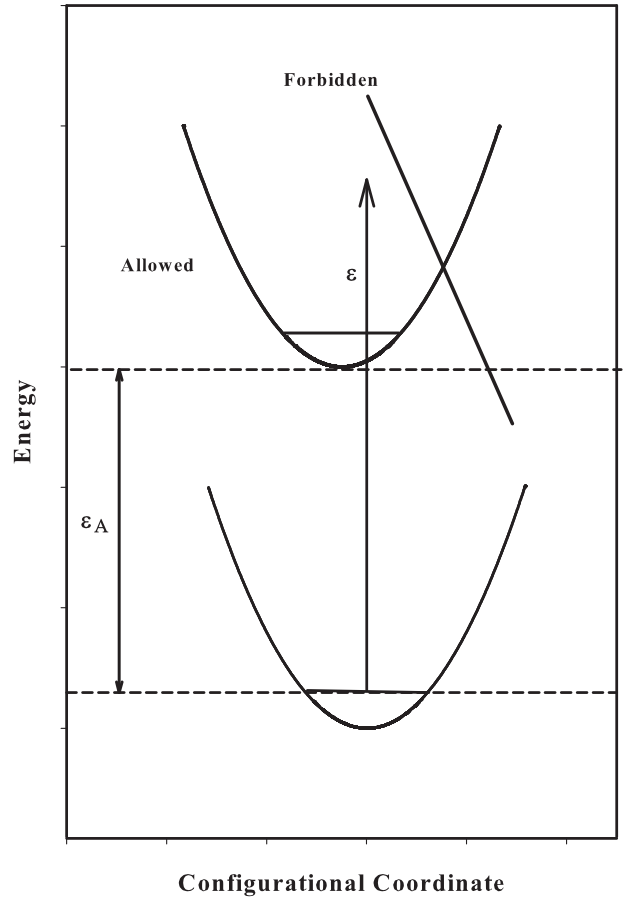,width=0.5\linewidth} 
\caption{Schematic diabatic potential energy curves
illustrating the model.} \label{Curveapply}
\end{figure}

\section{Electronic Absorption Spectra and Resonance Raman Excitation
Profile : predissociation of diatomic molecules} In this section we apply our method to the problem
involving Harmonic potential and linear potential. We consider a system of three
potential energy curves, ground electronic state and two
`crossing' excited electronic states (electronic transition to one
of them is assumed to be dipole forbidden and while it is allowed
to the other) \cite{AniThesis,AniBook,Ani1}. We calculate the electronic absorption spectra and resonance Raman
excitation profile to understand the effect of predissociation. The propagating wave functions on the excited
state potential energy curves are given by solution of the time
dependent Schr\"{o}dinger equation
\begin{equation}
\label{N33}i\frac \partial {\partial t}\left(
\begin{array}{c}
\psi _1^{vib}(x,t) \\
\psi _2^{vib}(x,t)
\end{array}
\right)=\left(
\begin{array}{cc}
H_{vib,e1}(x) & V_{12}(x) \\
V_{21}(x) & H_{vib,e2}(x)
\end{array}
\right) \left(
\begin{array}{c}
\psi _1^{vib}(x,t) \\
\psi _2^{vib}(x,t)
\end{array}
\right).
\end{equation}
\begin{figure} \centering
\epsfig{file=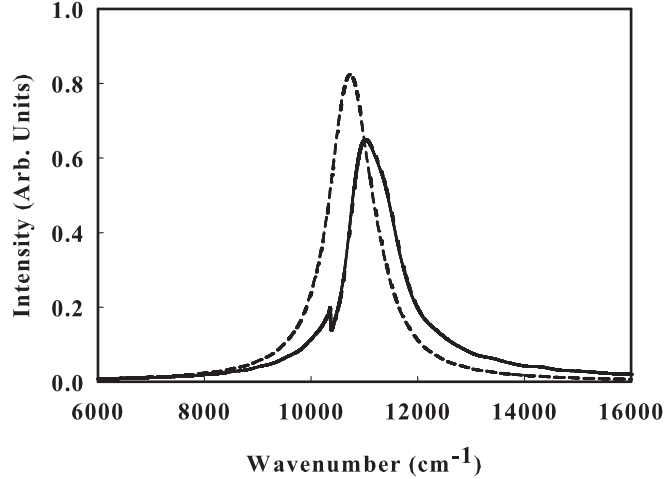,width=0.5\linewidth} 
\caption{Calculated electronic absorption spectra with coupling
(solid line) and without coupling (dashed line).} \label{Elec}
\end{figure}
In the above equation $H_{vib,e1}(x)$ and $H_{vib,e2}(x)$
describes the nuclear motion of the system in the first
electronic excited state (allowed) and second electronic excited
state (forbidden) respectively
\begin{equation}
\label{N34}H_{vib,e1}(x)=-\frac 1{2m}\frac{\partial ^2}{\partial
x^2}+\frac{1}{2}m \omega_{A}^2(x-a)^2
\end{equation}
and
\begin{equation}
\label{N35}H_{vib,e2}(x)=-\frac 1{2m}\frac{\partial ^2}{\partial
x^2}+K_{F}(x).
\end{equation}
In the above $m$ is the oscillator's mass, $\omega_{A}$ is the vibrational frequency on the
allowed states and $K_{F}$ is the slope of linear potential for the forbidden
states and $x$ is the nuclear coordinate. Shifts of the nuclear coordinate minimum upon excitation is given by 
$a$, and $V_{12}$ ($V_{21}$) represent coupling between the Harmonic and linear potential, which is taken to be
\begin{equation}
\label{N36}V_{21}(x)=V_{12}(x)=K_0\delta (x-x_c),
\end{equation}
where $K_0$ represent the strength of the coupling.
The intensity of electronic absorption spectra is given by
\cite{Zink,Heller}
\begin{eqnarray}
\label{N41}I_A(\omega )\propto & Re[\int_{-\infty }^\infty
dx\int_{-\infty }^\infty dx_0\Psi_i ^{vib*^{}}(x)\nonumber
\\ & iG(x,x_0;\omega +i\Gamma )\Psi_i ^{vib}(x_0)],
\end{eqnarray}
where
\begin{equation}
\label{N42}G(x,x_0;\omega +i\Gamma )=\langle x|[(\omega_0/2+\omega
-\omega _{eg})+i\Gamma -H_{vib,e}]^{-1}|x_0\rangle .
\end{equation}
and
\begin{equation}
\label{N42a} H_{vib,e}=\left(
\begin{array}{cc}
H_{vib,e1}(x) & K_0 |x_c\rangle\langle x_c| \\
K_0  |x_c\rangle\langle x_c| & H_{vib,e2}(x)
\end{array}
\right)
\end{equation}
Here, $\Gamma$ is a phenomenological damping constant which
account for the life time effects. $\Psi_i^{vib}(x,0)$ is given by
\begin{equation}
\label{N42b}\Psi _i^{vib}(x,0)=\left(
\begin{array}{c}
\chi_i(x) \\
0
\end{array}
\right),
\end{equation}
where $\chi_i(x)$ is the ground vibrational state of the ground
electronic state, $\omega_0$ is the vibrational frequency on the
ground electronic state, $\varepsilon_A$ is the energy difference
between the excited (allowed) and ground electronic state. Similarly resonance Raman scattering intensity can be expressed in
terms of Green's function and is given by \cite{Heller,Zink}.
\begin{figure} \centering
\epsfig{file=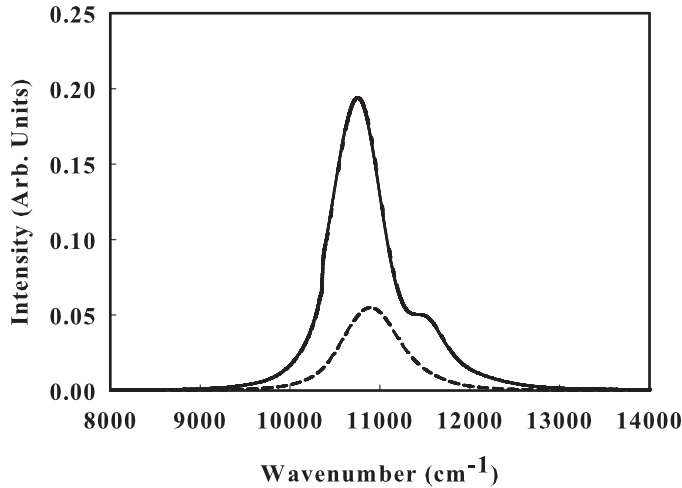,width=0.5\linewidth} 
\caption{Calculated resonance Raman excitation profile for excitation
from the ground vibrational state to the first excited
vibrational state, with coupling (solid line) and without coupling
(dashed line).} \label{Raman}
\end{figure}
\begin{eqnarray}
\label{N53} I_R(\omega )\propto &|\int_{-\infty }^\infty
dx\int_{-\infty
}^\infty dx_0\Psi _f^{vib*}(x,0)\nonumber\\
& iG(x,x_0;\omega+i\Gamma)\Psi _i^{vib}(x_0,0)|^2.
\end{eqnarray}
In the above $\Psi _f^{vib}(x,0)$ is given by
\begin{equation}
\label{N53a}\Psi _f^{vib}(x,0)=\left(
\begin{array}{c}
\chi _f(x) \\
0
\end{array}
\right),
\end{equation}
where $\chi _f(x)$ is the final vibrational state of the ground
electronic state. As $G_i^0(x,x_0;\omega)$ for the harmonic
potential as well as linear potential is known \cite{Grosche}, we can calculate
$G(x,x_0;\omega)$.

\subsection{Results using the model}
In the following we give results for the effect of curve crossing induced dissociation
on electronic absorption spectrum and resonance Raman excitation profile in the case where one dipole allowed electronic state
crosses with a dipole forbidden electronic state as in Fig.
\ref{Curveapply}. As in our earlier publications \cite{AniThesis,AniBook,Ani1}, here the ground state potential energy curve is
taken to be a harmonic one with its minimum at zero. The curve is constructed to be representative of the
potential energy along a metal-ligand stretching coordinate \cite{Zink}. We take the mass as $35.4$ amu and the vibrational wavenumber as
$400\:cm^{-1}$ \cite{Ani1} for the ground state. The first diabatic excited state potential energy curve is displaced by
$0.1\:\AA$ and is taken to have a vibrational wavenumber of $400\:cm^{-1}$. Transition to this state is allowed. The minimum
of the potential energy curve is taken to be above $10700\:cm^{-1}$ of that of the ground state curve. The second
diabatic excited state potential energy curve is taken to be a linear potential. Transition to this state is assumed to be dipole forbidden. The
two diabatic curves cross at an energy of $10804.1\:cm^{-1}$ with $x_c=0.02477\:\AA$. Value of $K_0$ we use in our calculation is
$K_0=5.54275\times 10^{-15}\:erg.\AA$. The lifetime of both the excited states are taken to be $450\:cm^{-1}$. The calculated
electronic absorption spectra is shown in Fig. \ref{Elec}. The profile shown by the dashed line is in the absence of any coupling
to the second potential energy curve. The full line has the effect of coupling in it. The calculated resonance Raman excitation
profile is shown in Fig. \ref{Raman}. The profile shown by the full line is calculated for the coupled potential energy curves.
The profile shown by the dashed line is calculated for the uncoupled potential energy curves. It is seen that curve crossing
effect can alter the absorption and Raman excitation profile significantly. However it is the Raman excitation profile that is
more effected. 
\section{Conclusions}
\noindent In our earlier papers we have proposed an exactly solvable model for the two state curve crossing problem \cite{AniThesis,AniBook,Ani1,Ani2,Ani3,Ani4,Ani5,AniRev}. In this paper we have extended our model to deal with the case of predissociation. We have analyzed the effect of curve crossing on electronic absorption spectrum and on Resonance Raman excitation profile for the case of Harmonic potential coupled to linear potential. We find that the Raman excitation profile is affected much more by the crossing than the electronic absorption spectrum \cite{AniThesis,AniBook,Ani1}. As we have shown, our model allows for an analytic solution, with a simple physical interpretation. The diabatic representation is the natural one for our model. On the other hand, the almost constant coupling leads to an adiabatic representation. We would therefore be inclined to conclude that in those cases where the diabatic picture is more convenient, our model should be more reliable than those in which we only have a slight breakdown of the Born-Oppenheimer approximation.

\section{Acknowledgments}
\noindent The author thanks Prof. K. L. Sebastian for continuous encouragements. It is a pleasure to thank Prof. M. S. Child for his kind interests, suggestions and encouragements.

\end{document}